\documentclass[12pt, preprint]{aastex}
\usepackage{epsfig}
\usepackage{natbib}

\def\ltsima{$\; \buildrel < \over \sim \;$}    
\def\lesssim{\lower.5ex\hbox{\ltsima}}           

\newcommand{\mHt}{\rm{H}_{2}}

\newcommand{\Rmol}{R_{\rm mol}}
\newcommand{\TCNM}{T_{CNM}}

\def\simless{\mathbin{\lower 3pt\hbox
   {$\rlap{\raise 5pt\hbox{$\char'074$}}\mathchar"7218$}}}   
\def\simgreat{\mathbin{\lower 3pt\hbox  
   {$\rlap{\raise 5pt\hbox{$\char'076$}}\mathchar"7218$}}} 

\begin{document}

\title{The Abundance of Molecular Hydrogen and its Correlation with
  Midplane Pressure in Galaxies: Non-Equilibrium, Turbulent, Chemical Models}

\author{Mordecai-Mark {Mac Low}$^{1,2}$  \& Simon~C.~O. Glover$^{2}$}
\affil{$^1$Department of Astrophysics, American Museum of Natural History, \\
       Central Park West at 79th Street, New York, NY 10024}
\affil{$^2$Zentrum der Astrophysik der Universit\"at Heidelberg,
  Institut f\"ur Theoretische Astrophysik, Albert-Ueberle-Stra{\ss}e
  2, 69120 Heidelberg, Germany}
\email{mordecai@amnh.org, glover@uni-heidelberg.de}

\begin{abstract}
  Observations of spiral galaxies show a strong linear correlation
  between the ratio of molecular to atomic hydrogen surface density
  $\Rmol$ and midplane pressure.  To explain this, we simulate
  three-dimensional, magnetized turbulence, including simplified
  treatments of non-equilibrium chemistry and the propagation of
  dissociating radiation, to follow the formation of $\mHt$ from cold
  atomic gas.  The formation time scale for $\mHt$ is sufficiently
  long that equilibrium is not reached within the 20--30~Myr lifetimes
  of molecular clouds.  The equilibrium balance between radiative
  dissociation and $\mHt$ formation on dust grains fails to predict
  the time-dependent molecular fractions we find.  A simple,
  time-dependent model of $\mHt$ formation can reproduce the gross
  behavior, although turbulent density perturbations increase
  molecular fractions by a factor of few above it.  In contradiction
  to equilibrium models, radiative dissociation of molecules plays
  little role in our model for diffuse radiation fields with strengths
  less than ten times that of the solar neighborhood, because of the
  effective self-shielding of $\mHt$.  The observed correlation of
  $\Rmol$ with pressure corresponds to a correlation with local gas
  density if the effective temperature in the cold neutral medium of
  galactic disks is roughly constant.  We indeed find such a
  correlation of $\Rmol$ with density. If we examine the value of
  $\Rmol$ in our local models after a free-fall time at their average
  density, as expected for models of molecular cloud formation by
  large-scale gravitational instability, our models reproduce the
  observed correlation over more than an order of magnitude range
  in density.
\end{abstract}


\section{Introduction}
\label{intro}

Stars are universally observed to form in molecular clouds.  Indeed,
recent observations have demonstrated that the
surface density of star formation $\Sigma_{\rm SFR}$ correlates
linearly with the surface density of molecular hydrogen
$\Sigma_{\mHt}$
\citep{rowndyoung99,wongblitz02,gaosolomon04,bigieletal08,bigieletal11}. This has led
to the suspicion that the formation of stars depends on the formation
of molecular hydrogen
\citep{schaye04,krumholzmckee05,elmegreen07,krumholzmckeetumlinson09b}.
However, the opposite view has also been argued: that molecular clouds
and then stars form from converging flows
\citep{ballesteroshartmann99,koyama00,hartmannetal01,vazquezetal06,ballesterosetal07,hennebelleetal08,heitsch08,inoue09},
primarily driven by large-scale gravitational instability
\citep{rafikov01,kim01,elmegreen02,kravtsov03,dalcanton04,maclowklessen04,li+05,li+06,shetty08,ostriker10}. 
In this picture, the formation of molecules is a consequence, not a cause, of the conditions 
required to form stars.

\citet{wongblitz02} and \citet{blitzrosolowsky04,blitzrosolowsky06}
added to the puzzle with the observation that in the inner parts of
spiral galaxies the ratio of molecular to atomic hydrogen surface density
\begin{equation} \Rmol \equiv \Sigma_{\mHt}/\Sigma_{\rm HI} \propto   
  P_m, \end{equation}
where $P_m$ is the midplane pressure and $\Sigma_{\rm HI}$
is the surface density of H~{\sc i}.  In detail, they derived an
empirical pressure dependence for $\Rmol$ that can be fit several
different ways.  Taking a common fit to all their individual data points yields
the fit \citep{blitzrosolowsky06}
\begin{equation} \label{eq:press}
\Rmol = \left[\frac{P_m / k}{(4.5 \pm 0.14) \times
    10^4}
    \right]^{0.94\pm 0.02},
\end{equation}  
where $P_m / k$ has units of K~cm$^{-3}$. 
This result was extended by \citet{leroy+08} who studied a broad
sample of spiral and dwarf galaxies from THINGS, and gave the results
in their Table~6. They found results quite consistent with
equation~(\ref{eq:press}). 

Both \citet{blitzrosolowsky06} and \citet{krumholzmckeetumlinson09a}
offered explanations of this observed
correlation that rely on the equilibrium balance
between radiative dissociation and H$_2$ molecule formation on the surfaces of
dust grains.  \citet{blitzrosolowsky06} built on the original suggestion by
\citet{elmegreen89} and \citet{elmegreenparravano94} that the
midplane pressure in disk galaxies determines molecular cloud properties.  They began
with the calculation by \citet{elmegreen93} that the equilibrium fraction of gas
in the molecular phase 
\begin{equation} f_{\rm mol} \equiv \Sigma_{\mHt} / \Sigma_{\rm tot} \end{equation} 
depends on both the interstellar pressure and the radiation field $j$ as
\begin{equation} \label{fmol} f_{\rm mol} \propto P_m^{2.2}
  j^{-1}. \end{equation} This equation was derived by assuming two
populations of spherical clouds, one with constant density,
representing diffuse clouds, and one with $r^{-2}$ density profiles,
representing hydrostatic, self-gravitating clouds.  Observations
\citep{heiles01} and turbulent models
\citep{avillez05,joung06,glover10} reveal that the distribution of
densities among the diffuse clouds is continuous rather than the delta
functions predicted by thermal phase transitions, demonstrating some
of the limitations of this approach.

 For low values of $\Rmol \lesssim 0.5$, the approximation $\Rmol
 \simeq f_{\rm mol}$ can be made.  \citet{blitzrosolowsky06} then
 invoked the observed correlation $\Sigma_{\rm SFR} \propto
 \Sigma_{\mHt}$, and assumed that the star formation rate determines
 the local radiation field directly ($j \propto \Sigma_{\rm SFR} $) to
 derive that
\begin{equation} \Rmol \propto P_m^{1.2}. \end{equation} However, the
discussion in \citet{blitzrosolowsky06} and \citet{leroy+08} neglects
that this relationship does not hold for the high values of $\Rmol \gg 0.5$
observed in some galaxies, as Equation~(\ref{fmol}) does not hold, nor
does the approximation $\Rmol \simeq f_{\rm mol}$.

\citet{krumholzmckeetumlinson09a,krumholzmckeetumlinson09b}, and
\citet{mckeekrumholz10} calculate the equilibrium value of $f_{\rm
  mol}$ in a uniform-density, spherical gas cloud exposed to far
ultraviolet (FUV) radiation. An essentially identical result was
already derived for a slab model of a cloud by \citet{sternberg88}.
They all demonstrate that the fraction depends on only two
parameters. The first parameter is the characteristic dust optical
depth $\tau_R = n_H \sigma_d R$ of a cloud of radius $R$, where $n_H$
is the number density of hydrogen nuclei in the atomic gas, and
$\sigma_d$ is the cross section for dust absorption per hydrogen atom.
The second parameter is the effective intensity of the ionizing
radiation
\begin{equation}
\chi = f_d \sigma_d c E_0^* / (n_H {\cal R}),
\end{equation}
(denoted $\alpha G$ by \citealt{sternberg88}), where $f_d$ is the
fraction of absorbed photons in the Lyman-Werner bands that result in
dissociation, $c E_0^*$ is the ambient flux of radiation in the
Lyman-Werner bands, and ${\cal R}$ is the rate coefficient for the
formation of H$_2$ on grains.

The effective intensity $\chi$ can be written in terms of the value of
$G_0'$, the ratio of $E_0^*$ to the typical dissociating radiation
field in the Milky way of $G_0 = 7.5 \times 10^{-4}$ photons~$^{-1}$
\citep{draine78}, as 
\begin{equation} \label{eq:chi}
\chi = 71 (\sigma_{d,-21} / {\cal R}_{-16.5}) (G_0^{'} / n_H),
\end{equation}
where $\sigma_{d,-21} = \sigma_d / 10^{-21} $~cm$^{-2}$ and ${\cal
  R}_{-16.5} = {\cal R} / 10^{-16.5}$~cm$^{3}$~s$^{-1}$.  Both scaled
constants are of order unity in the solar
neighborhood. \citet{krumholzmckeetumlinson09b} and
\citet{mckeekrumholz10} go on to express $\chi$ in terms of the
metallicity and the physical constants only, by assuming that the
atomic gas is in two-phase thermal equilibrium.  However, this second
step is not required.  Not taking it allows application of their
chemical model to a cloud of arbitrary density and ambient radiation
field.

These two papers then extend their approximation to compute $f_{\rm
  mol}$ in a spherical cloud composed of both atomic and molecular
gas, by making the assumption that the atomic gas is in pressure
equilibrium with the colder, denser molecular gas.  This allows them
to derive a fit to $f_{\rm mol}$ for the spherical gas cloud as a
function solely of $\chi$, $\Sigma_{\rm tot}$, the dust cross section
$\sigma_d$, and the mean mass per hydrogen nucleus $\mu_H$.  Using the
improved approximation given by \citet{mckeekrumholz10},
\begin{equation} \label{eq:kmt}
f_{\rm mol} \approx 1 +\left(\frac{3}{4} \right) \frac{s}{1+ s/4}, \end{equation} 
for $s<2$,  where,
\begin{equation}
s        = \ln (1 + 0.6\chi + 0.01 \chi^2)/(0.6 \tau_c), 
\end{equation}
and the optical depth through the atomic-molecular cloud $\tau_c =
0.75 \Sigma_{\rm tot} \sigma_d / \mu_H$.  For $s > 2$, $f_{\rm mol}
=0$ is a better approximation.  Essentially this same relationship was
used by \citet{krumholzmckeetumlinson09b} to derive the star formation
rate in galaxies, taking into account the observed correlation between
molecular hydrogen and star formation surface density.

But why should
H$_2$ exert such a strong influence on star formation?  Although H$_2$
is a coolant, it is effective only down to temperatures of 200~K,
whereas atomic fine structure emission (primarily from C$^{+}$) can
cool the gas down to temperatures of around 60~K in regions of low
dust extinction, or as low as 15~K in more highly shielded regions
where the photoelectric heating rate is small 
\citep{glovermaclow07b,gloverclark11}.  To cool
further, to the 10~K typical of most prestellar cores, generally
requires CO, which indeed forms efficiently only in H$_2$-rich
gas. However, the difference between 20~K and 10~K can only slightly
affect the star formation rate, as demonstrated by the recent simulations of
\citet{gloverclark11}.

On the other hand, gravitational instability produces dense gas that
quickly forms H$_2$ \citep{glovermaclow07b}.  Thus, H$_2$, and other
molecules that form with it, such as CO, may primarily just act to
trace dense gas that is already gravitationally unstable and
collapsing.  \citet{ostriker10} present an analytic model for how the
combination of dynamical and thermal equilibrium may lead to this
situation, also yielding a prediction of the correlation between the
molecular surface density and the midplane pressure based on the
linear relation derived between radiation field strength and pressure.
However, this model relies on the empirical correlation $\Sigma_{\rm
  SFR} \propto \Sigma_{\mHt}$ without offering a theoretical
explanation for it.

In this paper we examine the evolution of the molecular hydrogen to
atomic hydrogen ratio $\Rmol$ as a function of time and examine
whether that evolution can explain its observed correlation with
pressure, and thus its relationship to the star formation rate. To do
this, we have incorporated a simplified, non-equilibrium, chemical
network and radiative transfer model into a three-dimensional
simulation of supersonic (and super-Alfv\'enic), magnetohydrodynamical
turbulence. This allows us to follow the formation history of H$_2$,
starting from turbulent, magnetized atomic gas
\citep{glovermaclow07a,glover10}.

Our models follow the interplay between the cold neutral medium (CNM)
and molecular gas at different average densities.  They have velocity
dispersions of 5~km~s$^{-1}$, with resulting densities dispersed
around the mean value by more than four orders of magnitude \citep[see
Fig. 7 of ][]{glover10}.  Although we do not explicitly model
gravitational collapse, we do examine regions with varying average
density, providing a first approximation to the sequence of
quasi-static states that a large-scale, gravitationally contracting
region will pass through.  Only a small fraction of the molecular gas
will ever undergo full-scale gravitational collapse
\citep{goldreichkwan74,krumholztan07}, with the rest remaining at
pressures close to the surrounding atomic gas.  Because we do include
the full suite of heating and cooling processes relevant for both cold
atomic and molecular gas, our models offer insight into the behavior
of the cold gas even on large scales.
 
At any particular time after molecules begin forming, we find that the
molecular fraction $\Rmol$ varies primarily as a function of local gas
volume density, almost independently of radiation field strength,
except in diffuse regions with mean extinctions of $A_V \simless 0.3$.
This is because the strong self-shielding of H$_2$ means the
rate-limiting step in denser regions is the slow formation of
molecules on dust grains.  Significant mass fractions of the flow in
these models become fully molecular in only a few million years
\citep{glovermaclow07b}, but molecule formation continues over periods
exceeding 20~Myr, comparable to cloud lifetimes.  Thus, equilibrium
values of the molecular fraction cannot be relied on.  This is
contrary to the behavior of CO, whose fractional abundance is far more
sensitive to the dust column density (and thus radiation field
strength) than to the volume density, because of its much weaker
ability to self-shield \citep{glovermaclow10}.

If molecular clouds rarely or never reach equilibrium, then what
determines the observed molecular fraction?  The hypothesis that
global star formation is controlled by gravitational instability of
the gaseous disk has been considered at least since the calculation of
the gravitational instability criterion in such disks by
\citet{goldreich65}.  Numerical simulations by
\citet{kravtsov03,li+06}, and \citet{tasker06} supported this idea,
demonstrating that it naturally could explain observations of both
global \citep{kennicutt98} and local \citep{martin01,bigieletal08}
correlations between gas surface density and star formation rate.  The
clumpy nature of high-redshift galaxies can also be explained by this
hypothesis \citep{bournaud07,bournaud09}.  Molecular cloud formation
as a by product of gravitational instability has been argued to be a
natural consequence of this hypothesis
\citep{elmegreen02,maclowklessen04,li+05,ballesterosetal07,taskertan09}.
This suggests that the key timescale is the free-fall time
\begin{equation} \label{eq:tff}
t_{\rm ff} = (G \rho)^{-1/2},
\end{equation}
at the local average density $\rho = <\mu_H n_H>$. We show that our
non-equilibrium models indeed predict the observed molecular fractions
to occur after a free-fall time at the average density in each model.

In section~\ref{sec:numerics} we describe our numerical method and the
models that we ran.  We then offer a simple analytical model of
molecular hydrogen formation in section~\ref{sec:explanation} and
discuss how it succeeds and fails in comparison to the full numerical
results. In section~\ref{sec:comparisons} we compare our results to
equilibrium models and the observations, while in
section~\ref{sec:discussion} we offer caveats and consider the
implications of our results.

\section{Numerical Method}
\label{sec:numerics}

To study the formation of molecular hydrogen from atomic hydrogen, we
use mesoscale simulations of turbulent, magnetized, atomic gas
incorporating a non-equilibrium chemical network, including an
approximation to the background dissociating radiation field. 
In order to resolve the turbulent flow sufficiently well to follow the
formation of H$_2$, we neglect the large scale structure of the
interstellar medium, and instead simulate periodic boxes with side
length $L = 5$--20 pc having defined average initial number density $n_0$ (and other
parameters such as metallicity $Z$, turbulent velocity dispersion, and magnetic flux). 

We use a version of ZEUS-MP \citep{hayesetal06} modified to include a
subcycled chemical model and the associated radiative and chemical
heating and cooling.  The method was described in detail in
\citet{glover10} but for the models presented here we explore a wider
range of initial conditions and run several of the simulations for considerably longer.
Radiative transfer is included using a six-ray approximation, 
with column densities measured from the edges of the cube
\citep{glovermaclow07a}. Although this is not strictly self-consistent, it does capture
the basic dynamics of radiative dissociation well (also see
\citealt{glover10} and \citealt{glovermaclow10}). 

All of the models presented here include driven turbulence, rather
than the decaying turbulence used in \citet{glovermaclow07b}, with an
rms velocity $v_{rms} = 5 \mbox{ km s}^{-1}$, and an initial vertical
magnetic field of $B_0 = 5.85$~$\mu$G.  In Table~\ref{turb_runs} we
give the initial density $n_0$ and metallicity $Z$, the box size $L$,
the strength of the radiation field $G$ in units of $G_0$, and the
number of zones along one edge of the cube $nx$.

\section{Molecule Formation}
\label{sec:explanation}

\begin{figure}
\plotone{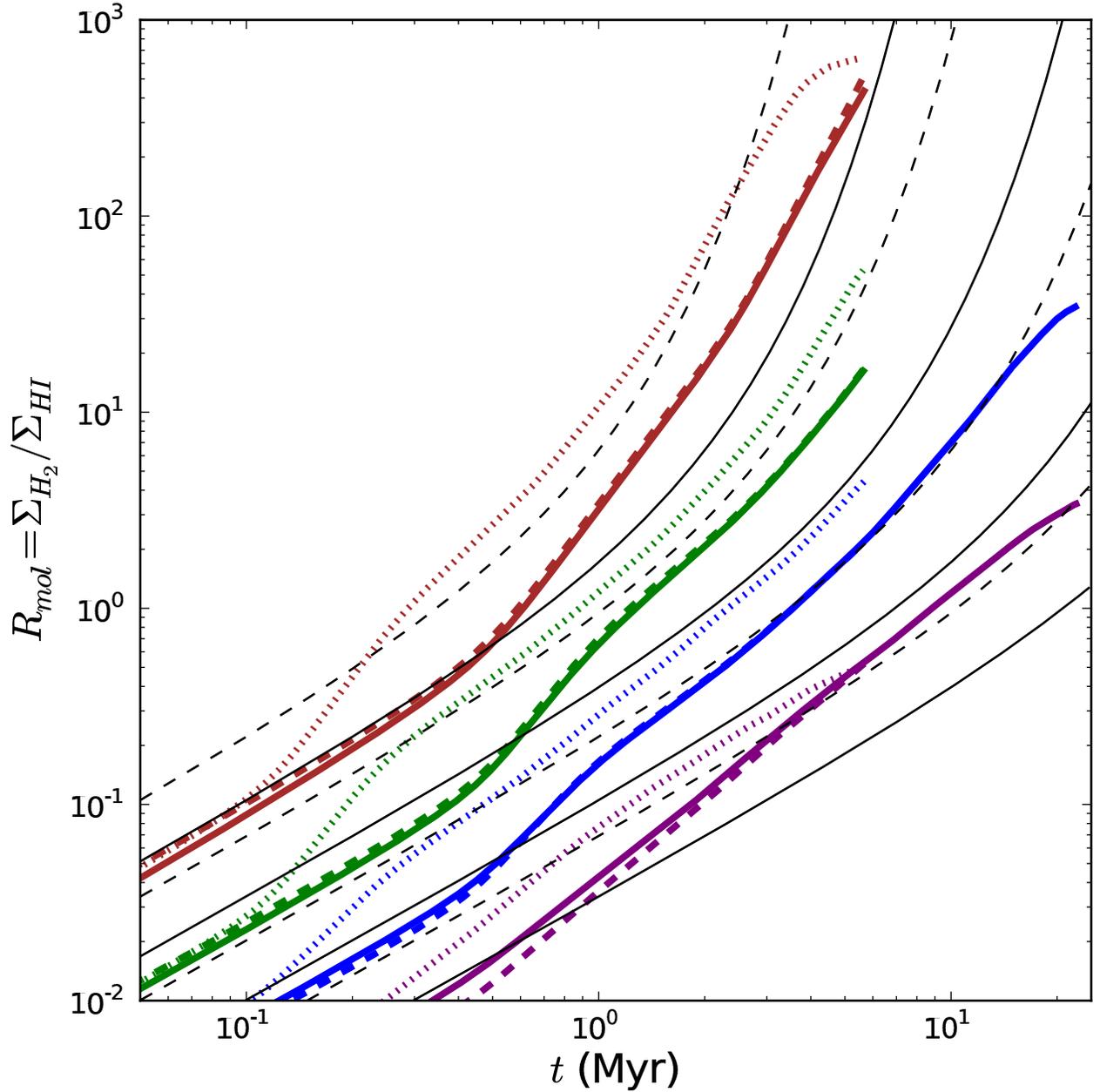}
\caption{\label{fig:result} Time history of the value of the column
  density ratio $\Rmol$ integrated through turbulent,
  magnetized boxes, starting with purely atomic gas, for models
  listed in Table~\ref{turb_runs}.  Models shown have average number
  density 30 {\em (purple, lowest}), 100 {\em (blue)}, 300 {\em (green)}, and
  1000 {\em (brown, highest)} cm$^{-3}$.  Canonical models {\em (solid)}
  are compared with G0 models lacking radiation {\em (dashed)}, and  -L5
  models with smaller boxes {\em (dotted)}.  The thin, solid, black
  curves show the analytic model given by
  Equation~(\ref{eq:analytic}) for time constants given by
  Equation~(\ref{eq:tauf}) for the densities of each model, while the
  thin, dashed, black curves show the analytic model with a clumping
  factor of two. The small box models diverge from the unclumped predictions
  earlier, as expected. No single clumping factor exactly reproduces
  all results.
}
\end{figure}

Our models show that the molecular to atomic ratio $\Rmol$ grows
almost as a power law in time over many dynamical times
(Figure~\ref{fig:result}).  Although our models will eventually reach
saturation, the time taken to do so is generally longer than the plausible
lifetimes for molecular clouds of 20--30~Myr \citep{fukui10}. We only
find evidence for reaching an equilibrium value of $\Rmol$ within the
first 25~Myr of evolution of our models for the model with the least
active molecule formation (the low density and metallicity model
n30-Z01 discussed in \S~\ref{subsec:equilibrium}). 

\subsection{Formation Time}
\label{subsec:formation}

We can derive a simple analytic model for the behavior seen in our
simulations by considering the finite formation timescale of molecular
hydrogen in a uniform medium. A rather similar approach is followed by
\citet{gnedin09}. The fractional local density of
molecular hydrogen
\begin{equation} X_{\mHt} = 2 n_{\mHt} / n, \end{equation} 
where $n = n_{\rm H} + 2 n_{\mHt}$, grows as 
\begin{equation}
\dot{X}_{\mHt} = C(t) n_{\rm H} {\cal R} (T),
\end{equation}
where the formation rate ${\cal R} \propto T^{1/2}$, and we have included a
time dependent clumping factor $C$ to account for the density perturbations.
We can eliminate $n_{\rm H}$ from these two equations to find
\begin{equation}
\dot{X}_{\mHt} = 2 C n_{\mHt} {\cal R}  (1 / X_{\mHt} - 1).
\end{equation}
Integrating this equation with a change of variable and the assumption
of a uniform value for the clumping factor $C$, we find
\begin{equation} \label{eq:analytic}
X_{\mHt} = 1 - \exp(- t / \tau_F), \end{equation}
where $\tau_F = 1 / (C {\cal R} n) $ is the time constant for $\mHt$ 
formation.
 \citet{hollenbach79} already showed that for typical
molecular cloud temperatures and without clumping ($C = 1$), 
\begin{equation} \label{eq:tauf}
\tau_F \simeq \frac{1 \mbox{ Gyr} }{n / (1 \mbox{ cm}^{-3})}. \end{equation}
The assumption of no clumping should be good for times much less than
the dynamical time $t_{dyn} = L/v_{rms}$, when the average density
rather than the peak clump density determines the evolution.  In our
canonical model, $t_{dyn} = 3.9$~Myr, while our small box models with
suffix L5 have $t_{dyn} = 0.98$~Myr.

We can compare this simple model with our numerical results and the
observations under the assumption that the local density ratio
$X_{\mHt}$ can be used to derive the column density ratio $\Rmol =
X_{\mHt} / (1 - X_{\mHt})$.  Figure~\ref{fig:result} shows that, as
expected, this analytic model works well at early times, up to roughly
0.4~Myr for the canonical model, and 0.1~Myr for our small box models.
In both cases, this represents a clumping time
\begin{equation} \label{eq:deltatc} \delta t_{c} \sim 0.1
  t_{dyn}. \end{equation}  Thereafter, $\Rmol$ increases above the
prediction of the simplest model, because of the enhanced clumping
caused by the turbulence \citep{glovermaclow07b}.

\subsection{Clumping}

The assumption of an equilibrium fraction of molecular hydrogen fails,
not just because of the long formation timescale of molecular hydrogen
but also because molecular hydrogen does not form uniformly.
Supersonic turbulence in the cold, neutral, atomic medium produces
strong, transient, density perturbations.  Peak densities exceed the
average value by the square of the local thermal Mach number, which
for conditions in the cold neutral medium can be a factor of $\delta n
/ n > 25$.  Since the formation timescale (Eq.~\ref{eq:analytic}) is
inversely proportional to density, formation proceeds far more quickly
in the density peaks.  Because of the strong self-shielding of $\mHt$,
it does not immediately photodissociate when it subsequently advects
into lower density regions, leading to non-equilibrium fractions in
lower density regions \citep{glovermaclow07b}.

No constant value of the clumping factor fits the late time results,
however.  For example, the value of $C = 2$ used in
Figure~\ref{fig:result} fits the $n = 30 \mbox{ cm}^{-3}$ results well
at late time, but overpredicts the canonical $n = 1000 \mbox{
  cm}^{-3}$ result substantially, though it does match the small box
model well at late times.  The time and density dependent variation of
the clumping factor has begun to be studied by
\citet{milosavljevic11}.

\section{Comparisons}

\label{sec:comparisons}
\subsection{Equilibrium Models}
\label{subsec:equilibrium}

\begin{figure}
\plotone{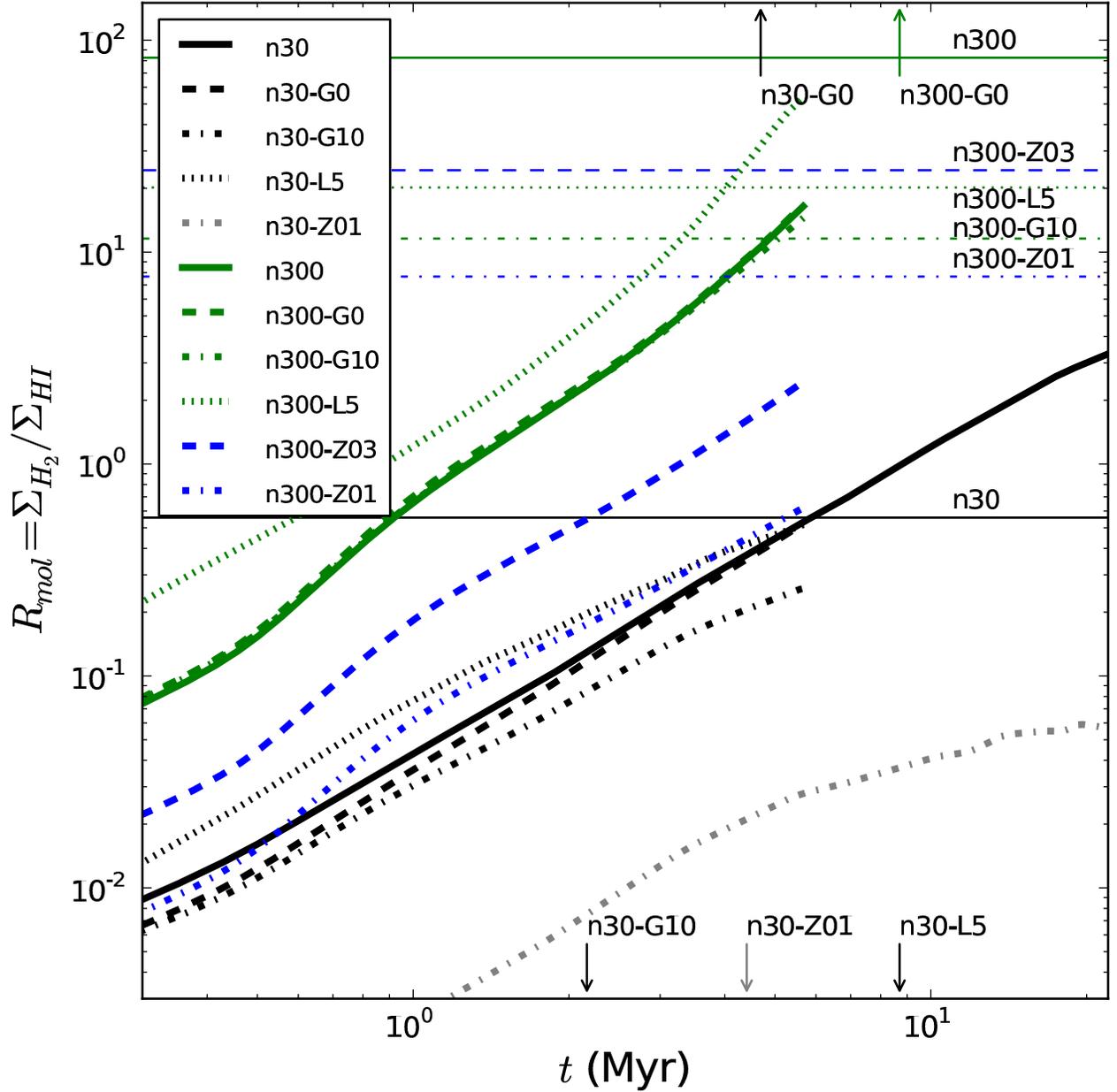}
\caption{\label{fig:kmt} Time history of $\Rmol$ for models with
  average number density 30 ({\em black, grey)}, and 300 {\em (green,
    blue)} cm$^{-3}$. Canonical models {\em (solid)} are compared with
  G0 models lacking radiation {\em (dashed)}, G10 models with ten
  times higher radiation {\em (dash-dotted)}, L5 models with smaller
  boxes {\em (dotted)}, and low metallicity models with 0.3 {\em (blue
    dot-dashed)} and 0.1 {\em (blue, grey dashed)} the canonical
  metallicity. The models with varying levels of radiation in the high
  density case almost entirely overlap each other. Thin horizontal
  lines and arrows show the predictions from the equilibrium
  calculation described by \citet{mckeekrumholz10} as given in
  Equation~(\ref{eq:kmt}). The predictions for the radiation-free G0
  models are $\Rmol \rightarrow \infty$, while the high-radiation,
  low-density case, as well as a number of other models, is predicted
  to have $\Rmol = 0$, as indicated by the arrows at the top and
  bottom of the plot.  The very weak dependence of our models on
  radiation strength directly contradicts the prediction of the
  equilibrium models.  }
\end{figure}

We compare a selection of our time-dependent results for turbulent gas
to the equilibrium values derived for spherical clouds by
\citet{mckeekrumholz10}.  In Figure~\ref{fig:kmt} we compare time
histories of a representative set of our models to the approximate
equilibrium values of $\Rmol$ for the conditions of each of these models given
by Equation~(\ref{eq:kmt}), taking $\Sigma_{\rm tot}$ to be the column
density through the cubical simulation domain, and $\Rmol = f_{\rm
  mol} / (1 - f_{\rm mol})$
The equilibrium approximation works particularly poorly at $\Rmol \ll
1$, predicting no molecules in many cases where we find measurable
molecular fractions.
At larger values of $\Rmol$, our models do intersect the spherical,
equilibrium values, but at widely varied times, and with no indication
of asymptotic approach to those values. 

The equilibrium models predict strong variation depending on the
ambient radiation field, while we see little to no effect (comparing
the G0 and G10 models to the canonical models, we see that they are
practically identical in the n300 case). This is because the
evolution is controlled by $\mHt$ formation rather than
photodissociation equilibrium.  This occurs because the supersonic
turbulent flows have strong density enhancements, leading both to
efficient self-shielding in the regions where most $\mHt$ is forming
and also fast local molecule growth in those enhancements.

We do note that the equilibrium models in the intermediate regime give
the qualitatively correct result of $\mHt$ formation at moderate
densities.  Although this is neither quantitatively corrrect nor
extendible to other regimes, it nevertheless means that observational
comparisons that rely only on this qualitative result can be
successfully made to the equilibrium models. 

The value of $\Rmol$ in our models depends on the metallicity, through
its regulation of the dust density, and thus the $\mHt$ formation
rate. However, only our very lowest density and metallicity model
n30-Z01 approaches an equilibrium value during the 20~Myr that we ran
our models.  This suggests that real molecular clouds rarely find
themselves in equilibrium with the diffuse radiation field because
their lifetimes are not long enough \citep{fukui10}.  Instead, they
are probably dissociated and dispersed by intense, local, UV radiation
from newly formed stars.

\subsection{Observations}
\label{subsec:observations}
Averaged over kiloparsec scales, the average temperature, magnetic
pressure, and turbulent velocity dispersion of the cold, neutral
medium vary little, with $\TCNM \sim 60$--70K
\citep{wolfire+95,wolfire03}, field strength $B \sim 5 \mu$G
\citep{heiles03} or perhaps 20\% higher as seen in observations of
external galaxies \citep{beck05}, and turbulent Mach number $M_{t}
\sim 5$ \citep{heiles03}. \citet{ostriker10} argue that this roughly
constant effective dynamical temperature comes from self-regulation of
star formation.  The roughly constant dynamical temperature suggests
that the apparent dependence of the fraction $\Rmol$ on pressure $P_m$
given in Equation~(\ref{eq:press}) may actually reflect a density
dependence, as already suggested by \citet{blitzrosolowsky04}.

Figure~\ref{fig:result} indeed shows a clear density dependence, but
it also shows strong time dependence, raising the question of what
molecular fraction should actually be compared with the observations.
If the hypothesis holds that molecular clouds actually form during
large-scale gravitational instability as discussed in \S~\ref{intro},
then the characteristic time in their lives at which clouds will be
observed is a free-fall time (Eq.~\ref{eq:tff}) after gravitational
collapse begins in cooled atomic gas.  Our models do not begin with
clumped atomic gas as the actual ISM does.  Instead, the turbulent
flows initialized on the grid clump the gas over a finite time $\delta
t_c$ (Eq.~\ref{eq:deltatc}).  In order to account for this effect, we
have taken the zero point in time for this comparison to be $\delta
t_c$ rather than the initial time of our models.  We emphasize that
the observations show an order of magnitude scatter around the average
$\Rmol$--$P_m$ correlation, so that using a characteristic time to
describe a time-dependent process appears likely to be consistent with
the observations if they catch clouds at different points in their
evolution centered around a free-fall time.

  Each of our models has a fixed average density $n_H$ given in
  Table~\ref{turb_runs}.  We use that value to compute the average
  pressure in our models
  \begin{equation}
  P = n_H k \TCNM + \langle B^2\rangle / 8 \pi + 0.5 \mu_H n_H v_{rms}^2,
  \end{equation}
  where we assume that $\TCNM = 60$~K, and the mean mass per particle
  $\mu_H = 2.11 \times 10^{-24}$ appropriate for atomic gas with 10\%
  helium by number.  Because our models are subject to dynamo
  generation of magnetic field from the initial value, we use the
  actual values of the average magnetic pressure at each time rather
  than the initial value. 

\begin{figure}
\plotone{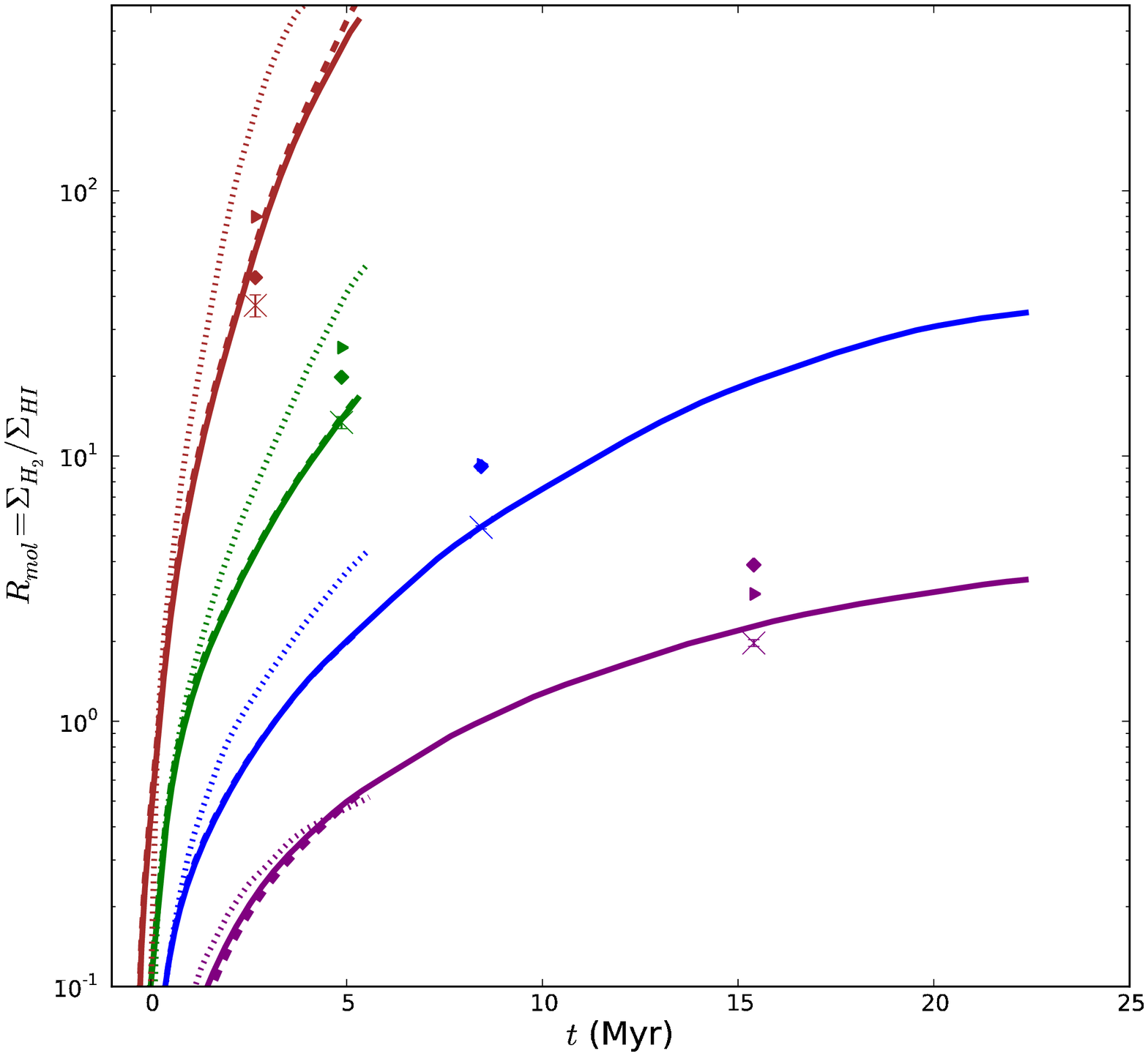}
\caption{\label{fig:H2frac} Time history of the value of the column
  density ratio $\Rmol$ for the same models presented in
  Figure~\ref{fig:result}, shown on semi-log axes for clarity.
  Superposed, we show the empirical fit for the canonical model of
  each density of \citet{blitzrosolowsky06}, including the errors {\em
    (X)}, as well as the spiral {\em (diamond)} and dwarf {\em
    (triangle)} galaxy samples of \citet{leroy+08}.  For all points we
  take a CNM temperature of 60~K, and the average density,
  time-dependent magnetic pressure, and turbulent pressure of each
  model.  They are placed on the figure at a free-fall time for the
  average density of that model, as described in the text.  To account
  for clumping of the atomic ISM, we use a zero time of $\delta t_c$
  in this plot, so that the initial time of each model is negative.  }
  \end{figure}

 We then use Equation~(\ref{eq:press}) to derive the
  value and standard deviation of $\Rmol$ predicted by
  \citet{blitzrosolowsky06} for the pressure for that model by setting
  $P_m = P$.  We also
  derive values of $\Rmol$ from the spiral and dwarf samples of
  \citet{leroy+08}.  The spiral sample has $ \Rmol = (10^{-4.23}
  P_m/k)^{0.80}$, while the dwarf sample has $\Rmol = (10^{-4.51}
  P_m/k)^{1.05}$. 
  All of these values are plotted in Figure~\ref{fig:H2frac} at times
  $t_{\rm ff} (n_H)$ after the beginning of evolution of our models.


  We find that our models predict the observed $\Rmol$ across more
  than an order of magnitude in pressure, with an accuracy better than
  the scatter in the observationally derived relations themselves.
  The ability to predict not just a single value, but a whole family of
  values accurately suggests that we have captured the
  important physics of the variation of the molecular fraction $\Rmol$
  with the estimated midplane pressure.

\section{Discussion}
\label{sec:discussion}

\subsection{Influence of numerical resolution}
\label{subsec:resn}

Figure~\ref{fig:H2frac} shows that the values of $\Rmol$ we find in
our $L = 5 \: {\rm pc}$ simulations are systematically higher at late
times than those found in our $L = 20 \: {\rm pc}$ simulations.   
(This is in addition to the early time differences discussed above
that occur because of the difference in crossing time $t_{\rm dyn}$
between the two box sizes.)  Since both sets of runs were performed
with the same numerical resolution of $nx = 128$ zones per side, the
spatial resolution of our L5 runs is four times larger than that in
our other runs. This prompts the concern that our simulations may be
under-resolved, and that our reported results for $\Rmol$ might have a
significant dependence on the spatial resolution of the simulations.

\begin{figure}[htb]
\epsscale{0.6}
\plotone{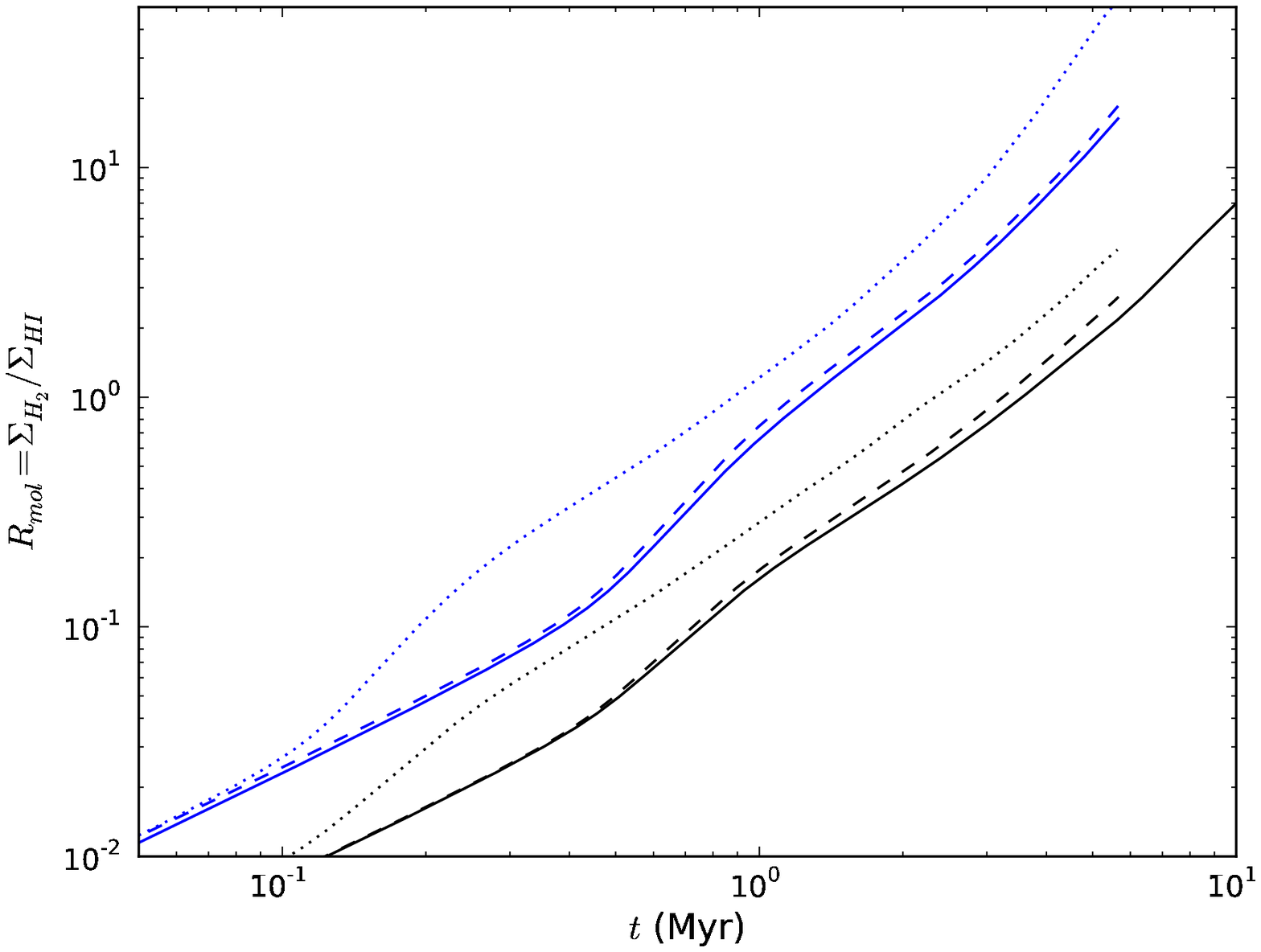}
\vspace{0.4in}
\plotone{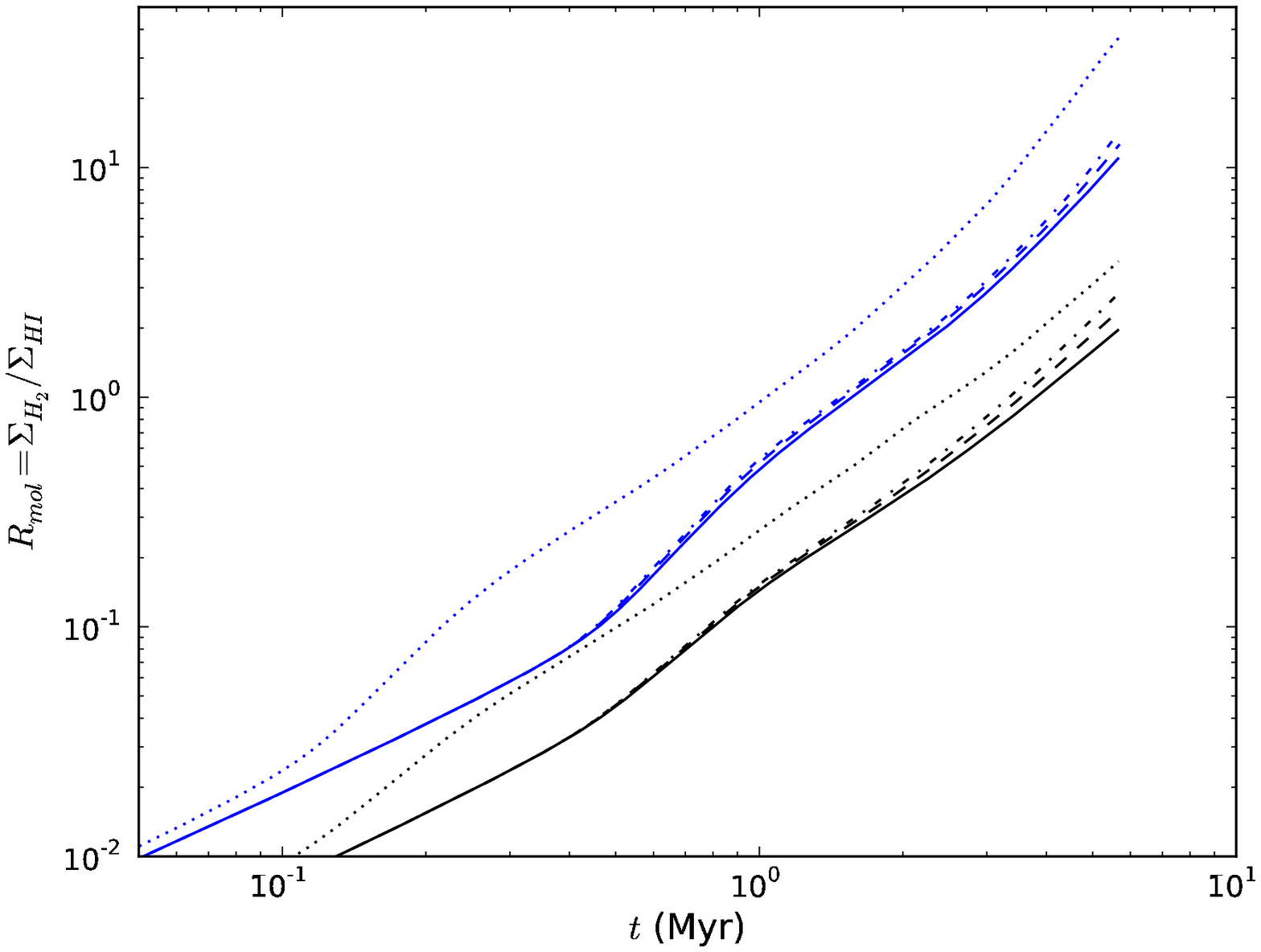}
\epsscale{1.0}
\caption{\label{fig:resn1} Resolution studies of the time history of
  $\Rmol$. (a) Comparison of runs n100 (black, solid line) and n300
  (blue, solid line) with
  runs n100-256 and n300-256 (dashed lines) having twice the numerical resolution,
  and with runs n100-L5 and n300-L5 (dotted lines), having four times the effective
  resolution but in a four times smaller box. (b) Resolution studies
  of the n100 (black) and n300 (blue) models reaching $512^3$ resolution by using the
  reduced chemical network of \citet{nl99}. Runs ending in NL-128
  (solid lines),  NL-256 (dashed lines), and NL-512 (dash-dotted
  lines) are directly compared to NL-L5 runs (dotted lines) that have physical
  resolution equivalent to the NL-512 runs. }
\end{figure}

To investigate this, we have performed several additional simulations
with a higher numerical resolution. In Figure~\ref{fig:resn1}(a), we
compare the evolution of $\Rmol$ in two canonical runs and the
corresponding L5 runs, with that in two runs performed using the same
sets of simulation parameters, but with twice the numerical
resolution. Increasing the resolution to $nx = 256$ zones per side
leads to slightly faster production of H$_{2}$ at late times,
particularly in the lower density runs, since we are better able to
resolve the highest density structures produced by the
turbulence. However, the effect is small, and does not explain the
significantly larger difference that we see between the results of the
5~pc (L5) and 20~pc runs.

Ideally, we would perform a similar comparison using results from runs
with $L = 20 \: {\rm pc}$ and a numerical resolution of $512^{3}$
zones, which would have the same spatial resolution as our $L = 5 \:
{\rm pc}$ runs. Unfortunately, the high computational cost of
performing a $512^{3}$ simulation using the full \citet{glover10}
cooling and chemistry model has so far rendered this impractical.

However, it has been possible to perform simulations with this
resolution by adopting a simpler treatment of the CO chemistry based
on \citet{nl99}. By using this simpler treatment, we accept a slightly
higher degree of error in our predicted CO, C and C$^{+}$ abundances,
and lose our ability to follow the evolution of chemical species such
as OH or water, but still retain the ability to follow the evolution
of the $\mHt$ abundance accurately, since our treatment of the
hydrogen chemistry remains the same.  At these densities, the change
in cooling from the modified chemistry will produce negligible changes
in the chemical evolution.  We show the results of this full
resolution study in Figure~\ref{fig:resn1}(b).  Once again, we see
that although our results remain somewhat sensitive to the numerical
resolution of the simulation, the effect is relatively small and
cannot be responsible for the difference between the $L = 5 \: {\rm
  pc}$ and $L = 20 \: {\rm pc}$ runs.

If spatial resolution is not the answer, then what is responsible for
the difference between the runs with different box sizes at late
times?  The lower visual extinction of the gas in the smaller box
allows stronger photoelectric heating, which leads to the gas in the
L5 runs having a systematically higher temperature than the gas in the
larger volume simulations. For example, the volume weighted mean
temperature of the gas at the end of run n300 is approximately 40~K,
while in run n300-L5 it is 66~K. This difference in temperatures leads
to a systematic difference in the H$_{2}$ formation rate coefficient,
owing to the $T^{1/2}$ temperature dependence of the rate coefficient,
and hence leads to a faster production of H$_{2}$ in the smaller
volume simulations.

\subsection{Caveats}

In this section, we discuss the limitations of this work.  The biggest problem is
that we are substituting a study of local isotropic turbulent regions
without explicit self-gravity for one of the dynamical interstellar
medium.  However, to reach the resolutions of less than 0.2
pc used here in any large-scale model of the interstellar medium
remains computationally challenging.  These resolutions are required
in order to resolve the turbulent density peaks that determine the
$\mHt$ formation rate. 

We approximate the external field of FUV dissociating radiation by
injecting radiation at the boundaries of our periodic box.  Although
this is physically inconsistent, we find that in this study it makes
virtually no difference, as models with and without FUV behave almost
identically. This is because the densities that we study are
sufficiently high that $\mHt$ self-shielding efficiently blocks
penetration of UV into the dense regions where $\mHt$ forms and
remains. This also means that the opening of voids in the density
distribution by the turbulence has much less effect on the $\mHt$
fraction than the creation of density peaks.

In our models, we do not include local sources of dissociating or
ionizing radiation.  Star formation of course produces such local
sources, which can be effective at dispersing molecular clouds on
short time scales. Turbulent models of such dispersal have not yet
been performed, but analytic models assuming spherical symmetry by
\citet{matzner02} and \citet{elmegreen07} show that large clouds can
be dispersed in a dynamical time.  Similarly, analytic models
including accretion show lifetimes comparable to the observed values
of around 20~Myr \citep{fukui10} in either a slab converging flow
\citep{zamora11} or a spherical geometry \citep{goldbaum11}.  These
relatively short cloud destruction time scales support our assumption
that most clouds are observed at an age of roughly a free-fall time.

Our chemical model has been simplified to allow computation on a
three-dimensional grid.  However, in the range of temperatures and
densities considered here, \citet{glover10} have shown that we
reproduce the results of the UMIST model to within a few tens of 
percent. Moreover, we do not expect any of the chemical processes
omitted from the model to have a significant impact on the fractional
abundance of molecular hydrogen, and hence our values for $\Rmol$
should be quite accurate.

Finally, we assume in this study that the dominant contributions to
the observed column densities of atomic and molecular hydrogen come
from cold, dense clouds, rather than from the warm, diffuse medium
surrounding these clouds. This is good to better than a factor of two
in the Milky Way where cold gas represents roughly $6 M_{\odot} \mbox{
  pc}^{-2}$ while warm atomic and ionized gas is a bit less than $5
M_{\odot} \mbox{ pc}^{-2}$ \citep{ferriere01}, and is probably better
in more massive disks, but will break down in outer disks or other
regions where the WNM component dominates the mass budget.  Since in
practice $\Rmol$ is only measured in regions where CO emission can be
detected, our approximation should be reasonable in the regime where
observations are available.

\subsection{Comparison to Other Tests}

\citet{krumholz10} tested the equilibrium models of
\citet{krumholzmckeetumlinson09b} (see eq.~\ref{eq:kmt}) against
relatively low-resolution (smallest zone size $\Delta x = 65$~pc)
global numerical simulations of $\mHt$ formation in galaxies.  The
simulations were described by \citet{gnedin10}, and used the $\mHt$
formation law of \citet{gnedin09}.  \citet{krumholz10} found good agreement in cases
where we find that our models reach high molecular fractions quickly (those
with high density and low radiation intensity), but poor agreement in
cases where we find that time-dependence matters. The formation law of
\citet{gnedin09} effectively reproduces the simple model that we have
described in Equation~(\ref{eq:analytic}).  The inability of global
models to resolve the density peaks is captured in the higher value of
the clumping factor they require to reproduce observations.  Because
lower mean densities are deduced, these models do still overestimate
the sensitivity of molecule formation to radiation.

\citet{fumagalli10} compared the equilibrium model given by
Equation~(\ref{eq:kmt}) to local (100~pc beam size) observations of the column density of
\ion{H}{1} and global (1~kpc beam size) observations of CO and
\ion{H}{1}. They demonstrate that the equilibrium model with freely
adjusted clumping factor predicts high
molecular fractions in regions of high local density, as we also
predict without such an adjustable factor, and that there is a metallicity dependence, again as we also
predict. 


\subsection{Implications for the Star Formation Rate in Galaxies}

We have demonstrated that the molecular fraction $\Rmol$ remains
strongly time dependent over periods of rather more than 20~Myr
(Fig.~\ref{fig:result}).  Our high-resolution numerical results
provide calibration for the time-dependent molecule formation model of
\citet{gnedin09} which assumes similar physics, but must rely on much
larger clumping factors to reproduce sub-grid scale
turbulence. Equilibrium values of $\Rmol$ are only reached at times
well beyond the lifetimes of real molecular clouds of 20--30~Myr
\citep{fukui10}. Therefore, explanations of the observed correlation
between midplane pressure and $\Rmol$ that rely on equilibrium values
seem unlikely to be correct. 

We find that the observed correlation can nevertheless be explained if
most molecular gas occurs in regions that have been forming $\mHt$ for
a free-fall time at their current number density
(Fig.~\ref{fig:H2frac}).  This is a natural prediction for the
gravitational instability model for star formation in galaxies.
Molecular cloud formation occurs as part of the same process of large
scale collapse that eventually forms stars, so the observed
correlation between molecular hydrogen and star formation rate surface
densities comes about because they have a common cause, not because
the first controls the second.

Our result appears almost independent of the strength of the far
ultraviolet dissociating radiation field during formation of the
molecular clouds, as can be seen by the close agreement of our
canonical models with otherwise identical models run without or with
stronger dissociating radiation.  In future work the upper limits of
this behavior must be examined, since clearly a strong enough
radiation field will eventually dissociate molecular gas, or at least
ionize enough of it to dynamically disperse it, allowing dissociation
to occur.  In particular, once massive star formation begins, the
intense local radiation field probably does dissociate or disperse the
remaining molecular hydrogen, terminating molecule formation
relatively quickly and determining the cloud lifetime.

The metallicity clearly influences the molecular hydrogen formation
rate in our models, through its regulation of the dust density
(Fig.~\ref{fig:kmt}).  The $\mHt$ formation rate depends on the
product $nZ$ of the number density and the metallicity
\citep{glovermaclow10} because of the role of dust grains in the
formation reaction, so this is expected.  However, we find a weaker
sensitivity than suggested by Equation~(\ref{eq:kmt}) derived from an
equilibrium model.  Detailed comparison of our results to galaxies
with varying metallicity should prove informative.

\acknowledgments We thank A. Leroy, E. Rosolowsky, \& L. Blitz for
insight into their data, and M. Krumholz, C. McKee, \& B. Elmegreen
for useful discussions. The clarity and emphasis of the revised
version benefited from comments by the anonymous referee and
M.-Y. Lee. M-MML was partly supported by NASA/SAO grant TM0-11008X, by
NASA/STScI grant HST-AR-11780.02-A, and by NSF grants AST 08-06558 and
AST 11-09395. SCOG acknowledges support from DFG grants KL1358/4
and KL1358/5, from a Heidelberg University Frontier Grant, funded as
part of the German Excellence Initiative, from the Bundesministerium
f\"ur Bildung und Forschung via the ASTRONET project STAR FORMAT
(grant 05A09VHA), and from the Baden-W\"urttemberg Stiftung via their
program International Collaboration II (grant PLS-SPII/18). The
numerical simulations discussed in this work were performed on the
{\em kolob} cluster at Heidelberg University, and on the {\em Ranger}
supercomputer at the Texas Advanced Computing Center, the latter under
NSF Teragrid allocation TG-MCA99S024.

\clearpage

\begin{deluxetable}{lrlrcc}
\tablecaption{Input parameters used for each simulation. 
\label{turb_runs}}
\tablewidth{0pt}
\tablehead{
\colhead{Run}  
& \colhead{$n_{\rm 0}$}  
& \colhead{$Z$} & \colhead{$L$} &  \colhead{$G$} & \colhead{$nx$}
\\
\colhead{ } & \colhead{(cm$^{-3}$)} & \colhead{($Z_{\odot}$)}  & \colhead{ (pc)} &
\colhead{($G_0$)} & \colhead{(zones)}
}
\startdata
n30          & 30 & 1 & 20  & 1   & 128 \\
n30-G0    & 30 & 1 & 20  & 0  & 128 \\
n30-G10    & 30 & 1 & 20  & 10  & 128 \\
n30-L5     & 30 & 1 & 5    & 1 & 128 \\
n30-Z01   & 30 & 0.1 & 20  & 1 & 128 \\
n100     & 100 & 1 & 20  & 1 & 128 \\
n100-G0           & 100 & 1 & 20  & 0 & 128 \\
n100-L5            & 100 & 1 & 5  & 1 & 128 \\
n100-Z01         & 100 & 0.1 & 20  & 1 & 128 \\
n100-256         & 100 & 1 & 20  & 1 & 256 \\
n100-NL-128    & 100 & 1 & 20  & 1 & 128 \\
n100-NL-256    & 100 & 1 & 20  & 1 & 256 \\
n100-NL-512    & 100 & 1 & 20  & 1 & 512 \\
n100-NL-L5  & 100 & 1 & 5  & 1 & 128 \\
n300    & 300 & 1 & 20  & 1 & 128 \\
n300-G0    & 300 & 1 & 20  & 0 & 128 \\
n300-G10    & 300 & 1 & 20  & 10 & 128 \\
n300-L5    & 300 & 1 & 5  & 1 & 128 \\
n300-Z01    & 300 & 0.1 & 20  & 1 & 128 \\
n300-Z03    & 300 & 0.3 & 20  & 1 & 128 \\
n300-256    & 300 & 1 & 20  & 1 & 256 \\
n300-NL-128    & 300 & 1 & 20  & 1 & 128 \\
n300-NL-256    & 300 & 1 & 20  & 1 & 256 \\
n300-NL-512    & 300 & 1 & 20  & 1 & 512 \\
n300-NL-L5    & 300 & 1 & 5  & 1 & 128 \\
n1000    & 1000 & 1 & 20  & 1 & 128 \\
n1000-G0    & 1000 & 1 & 20  & 0 & 128 \\
n1000-L5    & 1000 & 1 & 5  & 1 & 128 \\
\enddata
\tablecomments{$Z_{\odot}$ is the solar metallicity, while $G_0$ is the strength of the local interstellar
  radiation field as derived by \citet{draine78}.  NL models were run
  without full carbon chemistry (see section \ref{subsec:resn}).}
\end{deluxetable}

\end{document}